\documentclass[]{aastex631}

\usepackage{color}

\usepackage{multirow}
\submitjournal{ApJL}

\begin{document}

\title{The First Polarimetric View on Quasi-Periodic Oscillations in a Black Hole X-ray Binary}

\correspondingauthor{Lian Tao}
\email{taolian@ihep.ac.cn}
\author{Qing-Chang Zhao}
\affiliation{Key Laboratory for Particle Astrophysics, Institute of High Energy Physics, Chinese Academy of Sciences, 19B Yuquan Road, Beijing 100049, China} 
\affiliation{University of Chinese Academy of Sciences, Chinese Academy of Sciences, Beijing 100049, China}

\author{Lian Tao}
\affiliation{Key Laboratory for Particle Astrophysics, Institute of High Energy Physics, Chinese Academy of Sciences, 19B Yuquan Road, Beijing 100049, China} 

\author{Han-Cheng Li}
\affiliation{Department of Astronomy, University of Geneva, 16 Chemin d’Ecogia, Versoix, CH-1290, Switzerland}

\author{Shuang-Nan Zhang}
\affiliation{Key Laboratory for Particle Astrophysics, Institute of High Energy Physics, Chinese Academy of Sciences, 19B Yuquan Road, Beijing 100049, China} 

\author{Hua Feng}
\affiliation{Department of Astronomy, Tsinghua University, Beijing 100084, China}

\author{Ming-Yu Ge}
\affiliation{Key Laboratory for Particle Astrophysics, Institute of High Energy Physics, Chinese Academy of Sciences, 19B Yuquan Road, Beijing 100049, China} 

\author{Long Ji}
\affiliation{School of Physics and Astronomy, Sun Yat-sen University, Zhuhai, 519082, People’s Republic of China} 

\author{Ya-Nan Wang}
\affiliation{Key Laboratory of Optical Astronomy, National Astronomical Observatories, Chinese Academy of Sciences, Beĳing 100101, China}

\author{Yue Huang}
\affiliation{Key Laboratory for Particle Astrophysics, Institute of High Energy Physics, Chinese Academy of Sciences, 19B Yuquan Road, Beijing 100049, China} 

\author{Xiang Ma}
\affiliation{Key Laboratory for Particle Astrophysics, Institute of High Energy Physics, Chinese Academy of Sciences, 19B Yuquan Road, Beijing 100049, China} 

\author{Liang Zhang}
\affiliation{Key Laboratory for Particle Astrophysics, Institute of High Energy Physics, Chinese Academy of Sciences, 19B Yuquan Road, Beijing 100049, China} 

\author{Jin-Lu Qu}
\affiliation{Key Laboratory for Particle Astrophysics, Institute of High Energy Physics, Chinese Academy of Sciences, 19B Yuquan Road, Beijing 100049, China} 

\author{Yan-Jun Xu}
\affiliation{Key Laboratory for Particle Astrophysics, Institute of High Energy Physics, Chinese Academy of Sciences, 19B Yuquan Road, Beijing 100049, China} 

\author{Shu Zhang}
\affiliation{Key Laboratory for Particle Astrophysics, Institute of High Energy Physics, Chinese Academy of Sciences, 19B Yuquan Road, Beijing 100049, China} 

\author{Qian-Qing Yin}
\affiliation{Key Laboratory for Particle Astrophysics, Institute of High Energy Physics, Chinese Academy of Sciences, 19B Yuquan Road, Beijing 100049, China} 

\author{Qing-Cang Shui}
\affiliation{Key Laboratory for Particle Astrophysics, Institute of High Energy Physics, Chinese Academy of Sciences, 19B Yuquan Road, Beijing 100049, China} 
\affiliation{University of Chinese Academy of Sciences, Chinese Academy of Sciences, Beijing 100049, China} 

\author{Rui-Can Ma}
\affiliation{Key Laboratory for Particle Astrophysics, Institute of High Energy Physics, Chinese Academy of Sciences, 19B Yuquan Road, Beijing 100049, China} 

\author{Shu-Jie Zhao}
\affiliation{Key Laboratory for Particle Astrophysics, Institute of High Energy Physics, Chinese Academy of Sciences, 19B Yuquan Road, Beijing 100049, China} 

\author{Pan-Ping Li}
\affiliation{Key Laboratory for Particle Astrophysics, Institute of High Energy Physics, Chinese Academy of Sciences, 19B Yuquan Road, Beijing 100049, China} 

\author{Zi-Xu Yang}
\affiliation{School of Physics and Optoelectronic Engineering, Shandong Unversity of Technology, Zibo 255000, China} 

\author{He-Xin Liu}
\affiliation{Key Laboratory for Particle Astrophysics, Institute of High Energy Physics, Chinese Academy of Sciences, 19B Yuquan Road, Beijing 100049, China} 

\author{Wei Yu}
\affiliation{Key Laboratory for Particle Astrophysics, Institute of High Energy Physics, Chinese Academy of Sciences, 19B Yuquan Road, Beijing 100049, China} 




\begin{abstract}

We present the first polarimetric analysis of Quasi-Periodic Oscillations (QPO) in a black hole binary utilizing \textit{IXPE} data. Our study focuses on Swift J1727.8--1613, which experienced a massive outburst that was observed by various telescopes across different wavelengths. The \textit{IXPE} observation we studied was conducted during the Hard-Intermediate state. The polarization degree (PD) and polarization angle (PA) were measured at 4.28$\pm$0.20\% and $1.9^{\circ}\pm1.4^{\circ}$, respectively. Remarkably, significant QPO signals were detected during this observation, with a QPO frequency of approximately 1.34 Hz and a fractional root-mean-square (RMS) amplitude of about 12.3\%. Furthermore, we conducted a phase-resolved analysis of the QPO using the Hilbert-Huang transform technique. The photon index showed a strong modulation with respect to the QPO phase. In contrast, the PD and PA exhibit no modulations in relation to the QPO phase, which is inconsistent with the expectation of the Lense-Thirring precession of the inner flow. Further theoretical studies are needed to conform with the observational results.

\end{abstract}

\keywords{Accretion -- Polarimetry -- X-rays: binaries -- X-rays: individual (Swift J1727.8--1613)}


\section{Introduction} \label{sec:intro}

Most black hole binary systems are transients, spending most of their time in quiescence. Occasionally, they enter an outburst phase lasting for several weeks to months. For a complete outburst, they follow a ``q" track in the Hardness-Intensity Diagram (HID) \citep{Homan_hid,Fender_hid,Belloni_hid,Belloni2016}. During such an outburst, the source will experience several canonical accretion states: the low hard state (LHS), the intermediate state (IMS) and the high soft state (HHS) \citep{Belloni_hid,Belloni2016}. In the LHS, the spectrum is primarily dominated by the non-thermal component from the hot corona, with the thermal component from the disk being relatively weak, whereas in the HHS, the disk contribution becomes dominant. The IMS is the transition state from the LHS to the HSS. It can be further divided into the Hard-Intermediate State (HIMS) and the Soft-Intermediate State (SIMS). The spectrum of SIMS is softer compared to that of HIMS. In addition to differences in spectral shape, there are distinct characteristics in the timing variability features between the four canonical states.

The low-frequency Quasi-periodic oscillations (LFQPOs) are one of the most remarkable timing features \citep{van_der_klis_qpo}. They appear as a narrow peak in the power density spectrum (PDS) with frequencies ranging typically from 0.1 to 20\,Hz \citep[e.g.][]{Ingram_QPO}. The LFQPO can be further divided into three classes: Type-A, Type-B and Type-C based on their frequencies, root-mean-square (RMS) amplitude and quality factor \citep{Casella_qpo_classes}. The type-C QPO is often observed during the LHS and HIMS, while the type-B and type-A QPO typically appear in the SIMS and HSS, respectively. Since the type-C QPO is the most commonly seen and strongest, it has been extensively studied. However, the physical origin is still under debate. The models explaining its origin can be roughly divided into two categories. One class attributes QPO to geometric effects, with the Lense-Thirring precession model of the inner flow being one of the most popular in this category. Initially proposed by \citet{Stella_LT_model}, this model was further developed by \citet{Ingram_LT_model}, considering a precessing accretion flow within a truncated accretion disk \citep{Esin_truncated, Huang_truncated}. While successfully applied to some sources for explaining QPO properties, \citet{Nathan_GRS1915} recently discovered a very small truncated radius in GRS 1915+105, which does not match the prediction of this model. 

Another model in this category is the jet precession model, which has also been proposed to explain the large time lags observed in MAXI J1820+070 \citep{Ma_Jet_precession}. The other class of models explains QPO as an oscillation of an intrinsic property of the accretion flow, for example, in the \texttt{vkompth} model \citep{Karpouzas_vkompth, Bellavita_vkompth}. This model has been successfully applied to various sources, such as GRS 1915+105 \citep{Karpouzas_GRS1915_QPO, Garc_GRS1915, Mariano_1915}, MAXI J1535--671 \citep{Zhang_1535_QPO,Rawat_QPO_1535} and MAXI J1820+070 \citep{Marui_QPO_1820}, to fit the RMS and lag spectra of QPO. It was found, that the corona in the HIMS expands horizontally and has a size of a few 100 $R_{\rm g}$ \citep[e.g.][]{Marui_QPO_1820,Zhang_1535_QPO,Mariano_1915}, which questions the precession within the truncated disk.

Polarization measurements provide a different and crucial way for identifying QPO models. However, polarization measurements, especially the study of the evolution of polarization properties with the QPO phase, require the detection of a large number of photons, therefore necessitating that the black hole binary is in a very bright outburst and there is a sensitive X-ray polarimeter. The high brightness of a new X-ray transient, Swift J1727.8-1613, and its high-statistic Imaging X-ray Polarimeter Explorer (\textit{IXPE}) observations provide us with an opportunity to explore the polarization properties of QPO for the first time.

Swift J1727.8-1613 was first discovered with \textit{Swift}/BAT and was initially recognized as GRB 230824A \citep{Swift_discover_new_source}. 
This source underwent a giant outburst, detected with multi-wavelength telescopes, and the rapid flux increase supports it as a new Galactic X-ray transient \citep{Swift_identify_transient,Maxi_transient,Maxi_transient2}. With a peak flux of 7.6\,Crab (15--50 keV), Swift J1727.8-1613 is one of the brightest X-ray binary systems seen so far 
 \citep{Swift_QPO}. Optical observations have suggested that this source is a candidate black hole in a low-mass X-ray binary system \citep{Optical_confirm_black_hole}. 
Furthermore, radio detection conducted with the Very Large Array (VLA) measured a spectral index of $0.29\pm0.07$, consistent with the presence of a compact jet originating from a black hole X-ray binary in its hard X-ray spectral state \citep{Radio_jet}. \citet{Veledina_swiftj1727_ixpe} estimated the distance of this source at about 1.5\,kpc by comparing the flux of turning point of Swift J1727.8--1613 with those of GX 339--4 and MAXI J1820+070. Using the \textit{NICER} data, the spectral fitting indicates a spin of $\sim$0.99 and an inclination of $\sim$$48^{\circ}$ \citep{NICER_QPO}. \citet{IXPE_polarization} reported a significant polarization detection with a polarization degree (PD) of 4.71$\pm$0.27\% and polarization angle (PA) of 2.6$^{\circ}\pm$1.7$^{\circ}$, with the \textit{IXPE} observation in the 2--8 keV band. In addition, the PA in X-ray is aligned with those in sub-millimeter wavelengths, as measured by the Submillimeter Array (SMA) \citep{submili_polarization}, and is consistent with the optical polarization measurements \citep{optical_polarization} as well as the radio polarization measured by the Australia Telescope Compact Array (ATCA) \citep{Ingram_2023_1727}, indicating that the corona is extended in the disk plane in the LHS and HIMS \citep{Veledina_swiftj1727_ixpe,Ingram_2023_1727}.

During this outburst, significant QPOs were detected \citep{Swift_QPO,NICER_QPO,NICER_QPO2,AstroSat_QPO,AstroSat_QPO2,Veledina_swiftj1727_ixpe,mereminskiy2023hard,Ingram_2023_1727}. In this letter, we present a polarization analysis of QPOs, specifically focusing on the QPO phase-resolved results. This paper is organized as follows: We provide information about the observations and data reduction methods in Section \ref{sec:data_reduction}, present the results in Section \ref{sec:results}, and discuss the implications of the results in Section \ref{sec:disc_conclu}.

\section{Observations and Data Reduction} \label{sec:data_reduction}

\textit{IXPE}, a collaborative project between NASA and the Italian Space Agency, was launched into space aboard a Falcon 9 rocket on December 9, 2021. It is the first highly sensitive X-ray polarimeter operating in the 2--8 keV energy range \citep{Weisskopf_ixpe}. \textit{IXPE} is composed of three identical modules, each equipped with an X-ray grazing mirror and a polarization-sensitive detector unit (DU) \citep{Baldini_ixpe}. 

\begin{figure}
    \centering
    \includegraphics[width=0.9\linewidth]{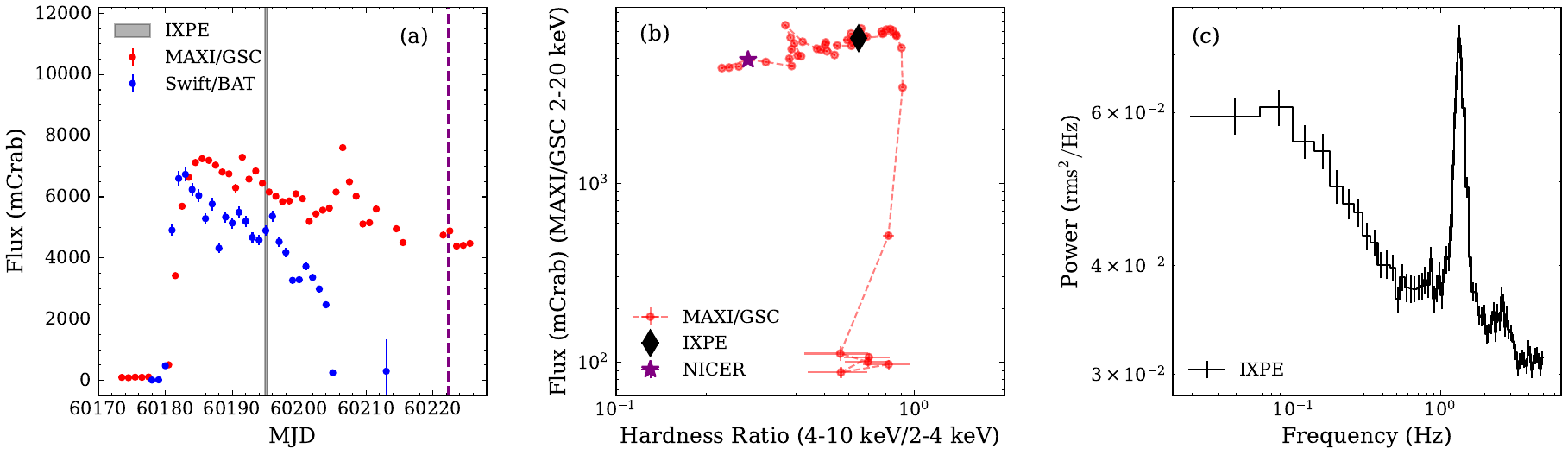}
    \caption{Panel (a): the light curves of MAXI/GSC (2--20 keV, red color) and Swift/BAT (15--50 keV, blue color) of Swift J1727.8--1613. The purple line represents the state transition time, and the gray strip represents the first observation of \textit{IXPE}. Panel (b): the hardness-intensity diagram (HID) of MAXI/GSC. The hardness ratio is defined as the count rate in the 4--10 keV band divided by that in the 2--4 keV band. The black diamond represents the location of \textit{IXPE}'s first observations. The purple star represents the location of state transition determined by the \textit{NICER} data \citep{NICER_SIMS}. Panel (c): the representative PDS from \textit{IXPE} (2--8 keV). The Poisson noise was not subtracted.}
    \label{fig:fig1}
\end{figure}

\textit{IXPE} conducted several observations for Swift J1727.8--1613. We analyzed the first observation (ObsID 02250901) when the QPO was strongest among all the \textit{IXPE} observations. The data was reducted using the publicly available software \texttt{ixpeobssim} version 30.6.3 \citep{Baldini_ixpeobssim}\footnote{\url{https://ixpeobssim.readthedocs.io/en/latest/}}. Our analysis started with the level 2 data products, which were downloaded from the HEASARC website\footnote{\url{https://heasarc.gsfc.nasa.gov/cgi-bin/W3Browse/w3browse.pl}}. To correct the photon arrival times for Solar System barycentric effects, we employed the \texttt{barycorr} tool within the \texttt{ftools}. A centroid circular region with a 96 arc-seconds radius was chosen as the source region. It is important to note that there is no need to subtract the negligible background after the source region selection given the high brightness of this source \citep{DiMarco_background}. We extracted events from the source region within the energy range of 2--8 keV using the \texttt{XPSELECT} tool within the \texttt{ixpeobssim} software. The observation was carried out within the time interval from 60194.83 to 60195.28 in MJD, as depicted in Figure~\ref{fig:fig1}. \citet{NICER_SIMS} reported a state transition from the HIMS to the SIMS or HSS on October 5 (MJD 60222) by investigating the timing properties, which is consistent with the observed radio quenching and subsequent radio flaring \citep{Radio_tran}. The transition location is marked by the purple line and star in panels (a) and (b) of Figure~\ref{fig:fig1}. Based on the position of the source in the HID, as shown in panel (b) of Figure~\ref{fig:fig1}, we can roughly categorize this observation as being in the HIMS. 

 This source exceeds 2\,Crabs during the observations, so a `gray' filter was placed in front of the detectors to reduce the incident flux \citep{Ferrazzoli_gray_filter,Soffitta_gray_filter,Veledina_swiftj1727_ixpe}. For the polarimetric analysis, \citet{Veledina_swiftj1727_ixpe} compared the \texttt{PCUBE} algorithm in \texttt{ixpeobssim} software and \texttt{XSPEC} \citep{Arnaud} method, concluding that the \texttt{XSPEC} approach is currently more accurate as it fully considers the corrections on instrumental response matrixes with the gray filter on. Therefore, we extract the stokes spectra with \texttt{PHA1, PHA1Q, PHA1U} algorithms in \texttt{ixpeobssim} with the \texttt{grayfilter} option set to \texttt{True}. These spectra are then fitted in XSPEC using the \texttt{Gain fit} command to circumvent the calibration issues when the `gray' filter is used.

\section{Analysis and Results} \label{sec:results}

\begin{table}
    \centering
    \begin{tabular}{ccccc} \hline
        Model & Parameter & \multicolumn{1}{c}{02250901}  \\ \hline
       Polconst  & PD (\%) & $4.28\pm0.20$\\
         & PA (deg) & $1.9\pm1.4$ \\  \hline
       Powerlaw  & $\Gamma$ & $1.85\pm0.01$   \\
         & Norm & $36.59^{+0.60}_{-0.46}$ \\ \hline
       Const  & Factor (DU1) & $1^{\rm fixed}$  \\
         &  Factor (DU2) & $0.977\pm0.001$  \\
         &  Factor (DU3) & $0.919\pm0.001$ \\ \hline
    Gain & Slope1 & $0.944\pm0.002$ \\
         & Offset1 ($10^{-2}$) & $5.40^{+0.38}_{-0.51}$  \\
         & Slope2 & $0.940\pm0.002$ \\
         & Offset2 ($10^{-2}$) & $5.17^{+0.49}_{-0.47}$ \\
         & Slope3 & $0.946\pm0.002$ \\
         & Offset3 ($10^{-2}$) & $8.27^{+0.44}_{-0.59}$ \\ \hline 
         & $\chi^{2}$/d.o.f & 1315.47/1329 \\ \hline 
                 
    \end{tabular}
    \caption{Spectro-polarimetric fitting results with  \texttt{Const*Polconst*Powerlaw} for Swift J1727.8--1613. Following \citet{Veledina_swiftj1727_ixpe}, the \texttt{gain fit} command in \texttt{XSPEC} is used to account for the calibration issues when the ‘gray’ filter is used}.
    \label{tab:tableA1}
\end{table}

\subsection{Spectro-polarimetric analysis} \label{subsec:dynamic_evolution}
We conducted fitting on the I, Q, and U Stokes spectra using  \texttt{Const*Polconst*Powerlaw} in \texttt{XSPEC}. The former model assumes constant energy dependence of PD and PA. The fitting results are listed in the Table~\ref{tab:tableA1}. All the uncertainties are given in a 68.3\% confidence level. In Figure~\ref{fig:figA1}, we display the I, Q, U stokes spectra and the best-model fit. We also tried the \texttt{Pollin} model, which assumes a linear energy dependence of PD and PA. The results estimated from the \texttt{Polconst} and \texttt{Pollin} models are roughly consistent. During the observation, QPO signals are significantly detected as shown in panel (c) of Figure~\ref{fig:fig1}. We extracted an average PDS in the 2--8 keV band, and respectively set the time resolution and the segment size to be 0.1 s (Nyquist frequency is 5 Hz) and 16 s by using the \texttt{POWSPEC} tool within \texttt{HEAsoft 6.32.0}. The logarithmic rebin is applied in frequency, where each bin size is $10^{\frac{3}{100}}$ times larger than the previous one. The PDS was normalized to the root-mean-squared per Hz \citep{Belloni_rms_norm,Miyamoto_norm}. Subsequently, the PDS was fitted with multiple Lorentzian functions plus a constant model \citep{Belloni_lore}. The fractional RMS amplitude of QPO is obtained by calculating the square root of the normalization of the Lorentzian function used to fit the QPO component, and the QPO frequency is the centroid frequency of this Lorentzian function. The QPO frequency and fractional RMS are approximately 1.34 Hz and 12.3\%, respectively.

\begin{figure}
    \centering
    \includegraphics[width=0.95\linewidth]{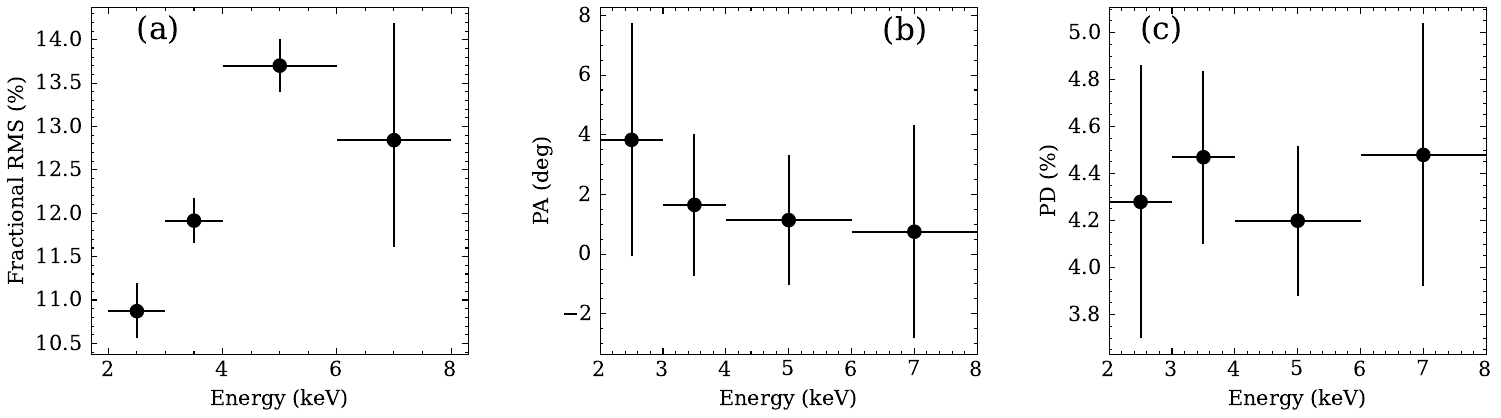}
    \caption{Panels (a)-(c): the energy dependence of QPO fractional RMS, PA and PD, measured from the \textit{IXPE} data of Swift J1727.8--1613.}
    \label{fig:fig2}
\end{figure}

To investigate the QPO properties and polarization evolution across different photon energy bands, we divided the data into multiple energy bins. We fitted the PDS as described in Section~\ref{sec:data_reduction}. The QPO frequency is consistent with being constant across different energies, while there is a noticeable positive trend in the fractional RMS of QPO with increasing photon energy, as shown in Figure~\ref{fig:fig2}. The last point shows relatively large uncertainties due to the low signal/noise ratio. In the energy-resolved analysis of polarization, due to the challenges in constraining spectra parameters within a very narrow energy band, we adopt the method of \citet{Veledina_swiftj1727_ixpe}. This involves fixing the spectral parameters in each sub-energy band to the values estimated from the 2–8 keV band using the \texttt{Const*Polconst*Powerlaw} model, and only allowing the PD and PA to vary. Since the \texttt{Polconst} model is used in the fitting, the PD and PA are treated as constant within each energy band. The fitted PAs and PDs both remain roughly unchanged with the photon energy considering the errors.

\subsection{Phase resolved analysis}
\begin{figure}
    \centering
    \includegraphics[width=0.75\linewidth]{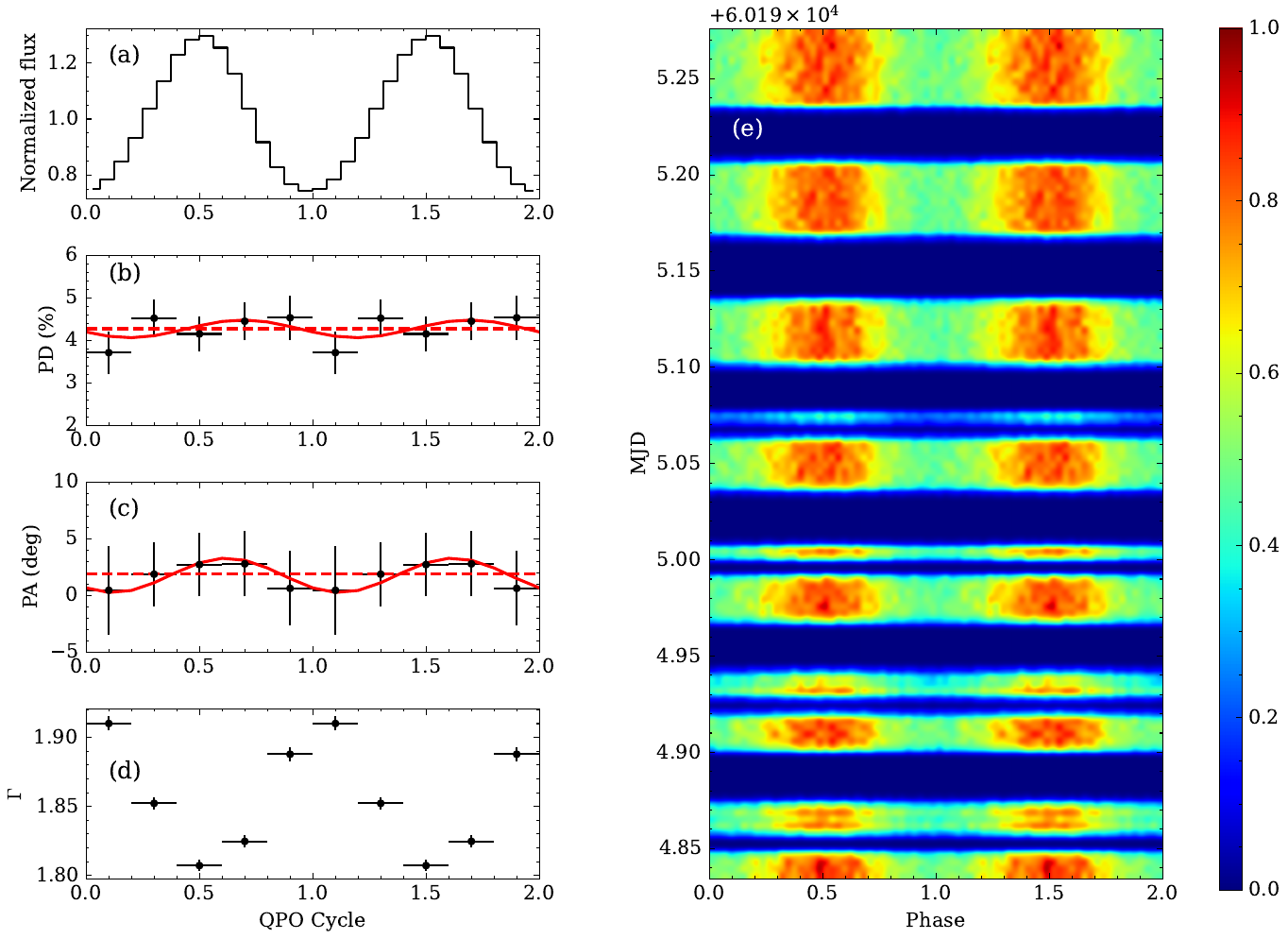}
    \caption{Panel (a): the phase-folded light curve on QPO period of Swift J1727.8--1613. The flux is normalized by averaged flux. Panel (b): the PD as a function of the QPO phase. The PD is estimated from the model \texttt{Const*Polconst*Powerlaw}. The horizontal dashed line represents the constant fit, while the solid line represents the sine function fit of PD. Panel (c): same as panel (b), but for PA. Panel (d): $\Gamma$ as a function of the QPO phase. Panel (e): the two-dimension QPO evolution with time.}
    \label{fig:fig3}
\end{figure}

We conducted a phase-resolved analysis using the Hilbert-Huang transform technique, which is a powerful tool for examining non-stationary signals \citep{Huang_HHT, Huang_HHT2}. This method has been previously applied in the phase-resolved analysis of QPOs \citep{Su_xtej1550_hht, Hsieh_mHz_hht, Yu_maxij1820_hht, Shui_maxij1820_hht}. Two essential steps are involved in this approach: mode decomposition and Hilbert transform analysis. Following the approach described in \citet{Shui_maxij1820_hht}, we employed Variational Mode Decomposition, which represents a more advanced method when compared to traditional decomposition techniques \citep{Dragomiretskiy_vmd}. To perform the mode decomposition, we utilized the open-source code \texttt{vmdpy v0.2}\footnote{\url{https://github.com/vrcarva/vmdpy}} \citep{VMDPY}. First, we extracted the light curves of each Good Time Intervals (GTIs) with 0.1\,s time resolution to perform the mode decomposition. For the segments with GTIs less than 100\,s, we have excluded them from the analysis. In the left panel of Figure~\ref{fig:figA2}, we display a representative 100-second-long light curve alongside the intrinsic mode functions (IMFs). Subsequently, we examine the PDS of the IMFs and compare it with the PDS of the original data. As depicted in the right panel of Figure~\ref{fig:figA2}, it becomes evident that the frequency of the second IMF (C2) is mostly close to the QPO frequency. Therefore we conducted the Hilbert-Huang transform on the second IMF to extract the instantaneous frequency and phase. By calculating the phase of the QPO, we were able to reconstruct the GTIs for different phases. The \texttt{maketime} command is used to construct the GTIs of different phase intervals. We divided five phase intervals, and events corresponding to these phase intervals were extracted using the \texttt{xselect} tool. Then we extracted the stokes spectra of each QPO phase interval using the \texttt{PHA1}, \texttt{PHA1Q}, \texttt{PHA1U} algorithms in \texttt{XPBIN}, and fitted them in \texttt{XSPEC} with \texttt{Const*Polconst*Powerlaw} model with the slope and offset of gain fixed to the values of Table~\ref{tab:tableA1}. The results are presented in Figure~\ref{fig:fig3}. As observed in panels (b) and (c), neither PD nor PA shows a significant trend with the QPO phase. We further fitted the evolution of PD and PA with phase using both constant and sine functions. The fitting results are summarized in Table~\ref{tab:table2}. Both constant and sine functions can provide a reasonable fit. Additionally, we observed a distinct modulation of photon index, $\Gamma$, with respect to the QPO phase. Specifically, $\Gamma$ reaches its minimum at the peak of the QPO and reaches its maximum at the valley of the QPO.

\begin{center}
\begin{table}
    \begin{tabular}{|c|c|ccccccc|}
        \hline
         & \multirow{3}{*}{Parameter} & \multicolumn{6}{c|}{QPO-Cycle dependence}                                                                                                                                                                                                                                                   \\ \cline{3-8} 
                                   &                            & \multicolumn{2}{c|}{Constant Fit}                                                                                & \multicolumn{4}{c|}{Sine Fit}                                                                                                                                                             \\ \cline{3-8} 
                                   &                            & \multicolumn{1}{c|}{$C$}                      & \multicolumn{1}{c|}{$\chi^2$/d.o.f} & \multicolumn{1}{c|}{$A$}                      & \multicolumn{1}{c|}{$B$}                      & \multicolumn{1}{c|}{$\phi_0$}               & \multicolumn{1}{c|}{$\chi^2$/d.o.f} \\ \hline
        \multirow{2}{*}{PolConst}          & PD (\%)                        & \multicolumn{1}{c|}{$4.3\pm0.2$} & \multicolumn{1}{c|}{2.06/4}         & \multicolumn{1}{c|}{$0.2\pm0.2$} & \multicolumn{1}{c|}{$4.3\pm0.2$}          & \multicolumn{1}{c|}{$0.40\pm0.3$} & \multicolumn{1}{c|}{1.86/2}         \\ \cline{2-8} 
                                   & PA ($^{\circ}$)                        & \multicolumn{1}{c|}{$1.8\pm1.4$} & \multicolumn{1}{c|}{0.47/4}         & \multicolumn{1}{c|}{$1.4^{+1.6}_{-1.0}$} & \multicolumn{1}{c|}{$1.8\pm1.4$} & \multicolumn{1}{c|}{$0.4^{+0.4}_{-0.2}$} & \multicolumn{1}{c|}{0.16/2}         \\ \hline
    \end{tabular}
    \caption{Constant fit and sine fit to the PD and PA variations with QPO phase in Swift J1727.8--1613. Constant function expression: $y = C$. The sine function expression: $y = A\sin[2\pi*(\phi-\phi_{0})]+B$. }
    \label{tab:table2}
\end{table}
\end{center}

\section{Discussion and Conclusion} \label{sec:disc_conclu}

In this study, we provide a polarimetric perspective on Swift J1727.8--1613. The remarkable brightness of this source, coupled with the presence of QPO, has offered us an opportunity to investigate the polarization properties of QPO. 

We classify this observation as being in the HIMS according to the location in the HID, and the spectrum is predominantly composed of Comptonized emission originating from the corona and the emission from the disk is weak (Liu et al. in preparation). The polarization measurements in the 2--8 keV energy range yield a PD of 4.28$\pm$0.20\% and a PA of 1.9$^{\circ}\pm$1.4$^{\circ}$, respectively. Notably, the relatively high PD is consistent with the 4.01$\pm$0.20\% detection observed in Cygnus X-1 during the hard state \citep{Cyg_X-1_polarization_IXPE_team, Cyg_X-1_polarization}. The observed PA in X-ray band aligns with the polarization measurements at optical, sub-millimeter, and radio wavelengths \citep{submili_polarization, optical_polarization,Ingram_2023_1727}. Therefore, assuming that the PA at radio and submillimeter wavelengths aligns with the direction of the jet is consistent with the notion that the corona extends perpendicular to the jet axis, i.e., parallel to the accretion disk \citep{Veledina_swiftj1727_ixpe,Ingram_2023_1727}. This alignment arises because multiple scatterings within the plane of the corona tend to polarize the X-rays in a direction perpendicular to that plane, different from the case of Sco X--1 \citep{Long_sco_X-1}. As a result, the coronal geometry of Swift J1727.8--1613 could potentially conform to models such as the `sandwich corona' \citep{Haardt_slab} or a hot inner flow within a truncated disk model \citep{Meyer_truncated_model}. Additionally, the PD and PA do not exhibit a significant evolution trend with photon energy (Figure~\ref{fig:fig2}), which further closely agrees with the predictions of the hot inner flow within a truncated disk, as depicted in Figure S9 of \citet{Cyg_X-1_polarization_IXPE_team} and Figure 5 of \citet{2023ApJ...949L..10P}.

Significant QPO signals are detected in this source, with a QPO frequency of $\sim$1.34 Hz and a fractional RMS amplitude of about 12.3\%. As discussed above, the emission is dominated by the corona, and the coronal geometry extends along the plane of the disk. Therefore, we first test the model proposed by \citet{Ingram_LT_model}, which involves a precessing accretion flow within a truncated accretion disk to explain the properties of QPO. As shown in Figure~\ref{fig:fig2}, the fractional RMS increases with photon energy in the 2--8 keV range, which is consistent with previous studies in other BH systems \citep{Zhang_GX339_4_QPO, Huang_MAXIJ1535_QPO, 
Belloni_1348_QPO, Zhang_GRS1915_QPO, Karpouzas_GRS1915_QPO, Ma_Jet_precession, Marui_QPO_1820}, and agree with the hypothesis that the QPO originates from the inner hot flow. In addition, as illustrated in Figure~\ref{fig:fig3}, $\Gamma$ exhibits a distinct modulation with the QPO phase. Specifically, $\Gamma$ reaches its minimum at the peak of the flux. Based on the scenario of Lense-Thirring precession, if we consider the evolution of flux as the projection effect, when the inner region becomes more visible, the projection area is at its maximum, and thus the flux is at its maximum. Meanwhile, high-energy photons from the inner region contribute more to the emission. As a result, the spectrum appears to be harder at the peak of the flux.

Furthermore, this model also predicts polarization modulation as a function of QPO phase \citep{Ingram_QPO_polarization_modulation}. They considered a truncated radius of 20 gravitational radii ($R_{\rm g}$) in the simulations. In the framework of the Lense-Thirring precession of a hot inner flow, a QPO of $\sim$1.34 Hz detected in Swift J1727.8--1613 approximately corresponds to a truncated radius of $\sim$ 15\,$R_{\rm g}$, by assuming that $M_{\rm BH}$ = 10 $M_\odot$ and $a = 0.98$.
This radius is essentially comparable to the radius of 20\,$R_{\rm g}$, allowing us to directly compare our results with the simulations. As shown in Figure~\ref{fig:fig3} and Table~\ref{tab:table2}, we find that both the PD and PA exhibit no modulations with respect to the QPO phase. Additionally, the polarization modulation is highly dependent on the viewing angle. A moderate inclination in this source is supported by spectral fitting results \citep[Peng et al. submitted, Liu et al. in preparation] {NICER_QPO}, and is consistent with the inclination estimated from the averaged PD \citep{Veledina_swiftj1727_ixpe}. If we assume a moderate inclination (30--60$^{\circ}$), the PD and PA modulation amplitudes from the model predictions should be higher than $\sim 0.5$\% and $\sim 4^{\circ}$, respectively \citep{Ingram_QPO_polarization_modulation}, significantly larger than the observed values of $0.2\pm0.2$\% and $1.4^{+1.6}_{-1.0}$$^{\circ}$. In a high inclination scenario, weak modulation amplitudes might be expected since the photons from the back of the inner flow might be observed. When these photons pass close to the black hole, depolarization will occur and the PA may change because of the General Relativistic (GR) effects, e.g., light bending. However, a high inclination would result in a higher PD ($\geq$ 7\%) than the observed value of $\sim$4\%.   

As mentioned in the introduction (Section~\ref{sec:intro}), there are other models proposed to explain the properties of QPO, for instance, the \texttt{vkompth} model. This model has been successfully applied to several sources to explain the energy dependence of the RMS and phase lag of QPO. The RMS energy dependence shown in Figure~\ref{fig:fig2} can also be explained by this model, since assuming constant $\tau$ and modulated $kT_e$, the modulation of $\Gamma$ would also be expected. However, since this model does not predict the modulation of the polarization with the QPO phase, we can not directly compare the observations with the model predictions.

In summary, we have provided the first polarimetric analysis of the QPO in Swift J1727.8--1613. Our findings indicate that both the PD and PA show no modulations with respect to the QPO phase. To gain a deeper understanding of the underlying physics, it is essential to conduct further theoretical studies in the future. Moreover, in order to better discern the polarization modulation in conjunction with the QPO phase, it is necessary to acquire simultaneous polarization, high-statistic timing and broad-band spectroscopy data. This can be achieved through an upcoming mission, the enhanced X-ray Timing and Polarimetry (eXTP) observatory \citep{Zhang_eXTP}.

\begin{acknowledgments}
We thank the anonymous referee for insightful comments and suggestions, which allowed us to improve and clarify the manuscript. Q. C. Zhao thanks Fabio Muleri and Alexandra Veledina for clarifying questions about the \texttt{PCUBE} and \texttt{XSPEC} analyses with the gray filter on. 
This study has made use of data sourced from the High Energy Astrophysics Science Archive Research Center Online Service, provided by the NASA/Goddard Space Flight Center. This work is supported by the National Key R\&D Program of China (2021YFA0718500). We acknowledge funding support from the National Natural Science Foundation of China (NSFC) under grant No. 12122306, 12333007 and 12027803, the CAS Pioneer Hundred Talent Program Y8291130K2, the Scientific and technological innovation project of IHEP Y7515570U1, and International Partnership Program of Chinese Academy of Sciences under grant No.113111KYSB20190020. H. C. Li acknowledge the support of the Swiss National Science Foundation.
\end{acknowledgments}

%

\vspace{5mm}





\appendix
\counterwithin{table}{section}
\counterwithin{figure}{section}
\section{Supplementary material}
\begin{figure}
    \centering
    \includegraphics[width=0.85\linewidth]{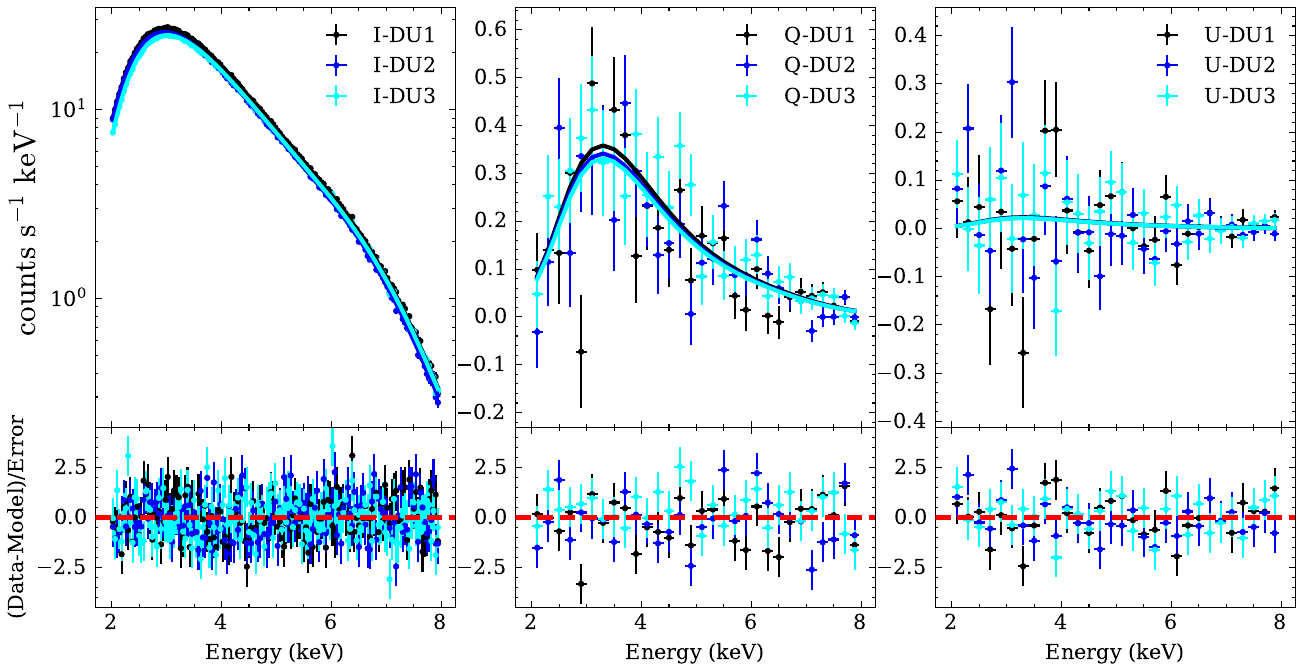}
    \caption{Left, Middle and Right panels: the stokes spectra of I, Q, U for ObsID 02250901 of Swift J1727.8--1613. The solid lines represent the best fit of \texttt{Const*Polconst*Powerlaw}.}
    \label{fig:figA1}
\end{figure}

\begin{figure}
    \centering
    \includegraphics[width=0.45\linewidth]{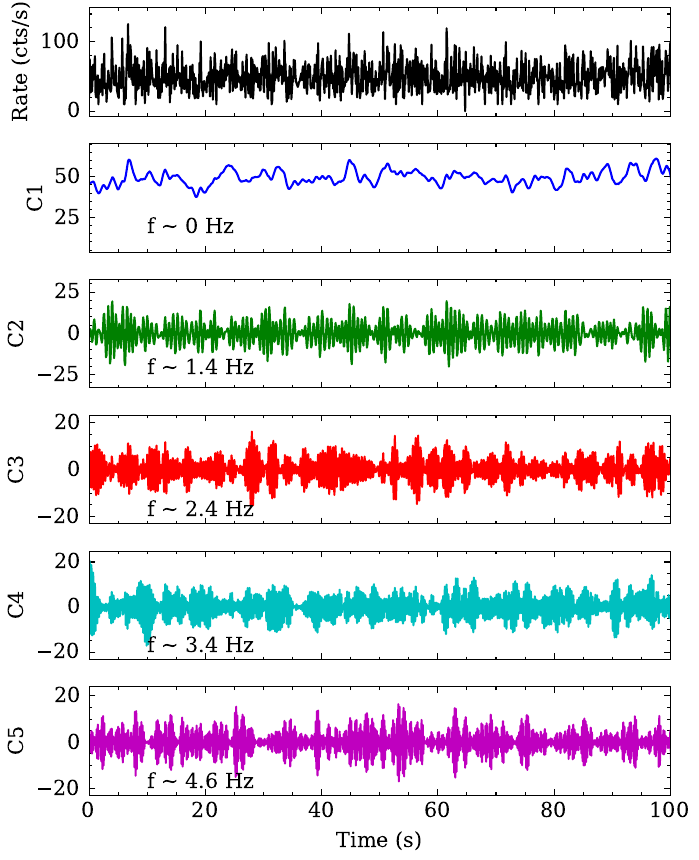}
    \includegraphics[width=0.455\linewidth]{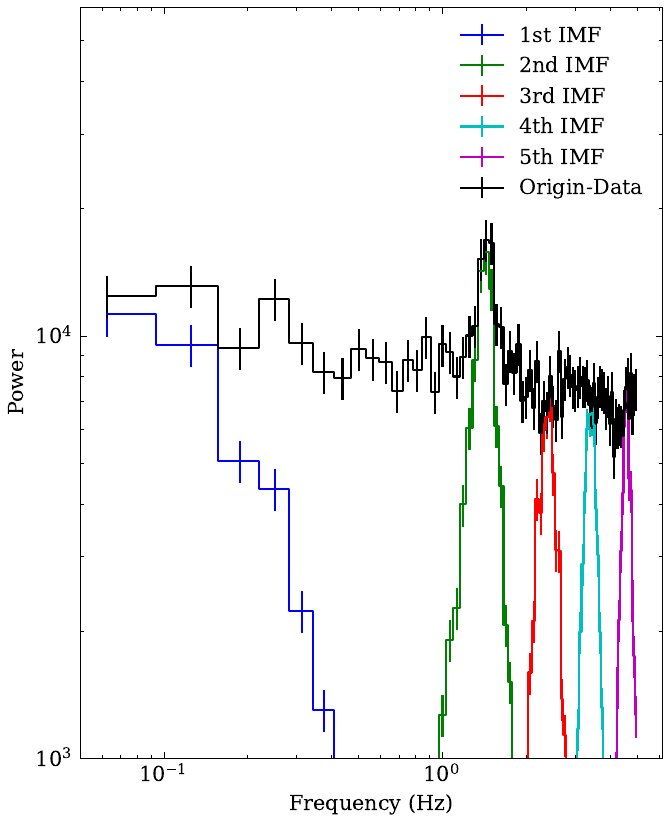}
    \caption{Left panel: representative example of a 100\,s light curve of Swift J1727.8--1613 and the corresponding IMFs (C1-C5). Right panel: the PDS of the original light curve and the 1st to 5th IMFs.}
    \label{fig:figA2}
\end{figure}

\bibliography{SwiftJ1727}{}
\bibliographystyle{aasjournal}



\end{document}